# Model-Based User Interface Design for Generating E-Forms in the Context of an E-Government Project

Bedir Tekinerdogan [1], Namik Aktekin[2]

[1] Wageningen University, Information Technology, Wageningen, The Netherlands
bedir.tekinerdogan@wur.nl

[2] Excellence Group, Hengelo, P.O. Box 768, 7550 AT, Hengelo, The Netherlands
n.aktekin@exxellence.nl

**Abstract.** We report on our experiences in an e-government project for supporting the automatic generation of E-forms for services provided by local governments. The approach requires the integration of both the model-based user interface design (MBUID) and software product line engineering approaches. During the domain engineering activity the commonality and variability of product services is modeled using feature diagrams and the corresponding UI models are defined. To support the automation of e-forms the implemented feature models are on their turn used to generate E-forms automatically to enhance productivity, increase quality and reduce cost of development. We have developed three different approaches for e-form generation in increasing complexity: (1) offline model transformation without interaction (2) model transformation with initial interaction (3) model-transformation with run-time interaction. We discuss the lessons learned and propose a systematic approach for defining model transformations that is based on an interactive paradigm.

## 1. Introduction

Over the last decade, the modernization of government administrations using Information and Communications Technology (ICT) has culminated in the definition of so-called e-government [2][19][13]. E-government is a general term describing the use of ICT to facilitate the operation of government to provide better public services to citizens and businesses. The target of the services can be different and as such e-government includes different models including *government-to-government (G2G)*, *government-to-business (G2B)*, and *government-to-citizen (G2C)*. The context of this study is an industrial e-government project which aims to use ICT to provide and improve services of local governments. The research has been carried out together with Exxellence, a medium-size ICT company in Enschede, The Netherlands [16]. In the context of the project, we have focused on the model of *local government-to-citizen* which aims to support the interaction between local and central government and private individuals. One of the objectives of Exxellence is the generation of electronic forms (e-forms) for supporting administrative tasks of different local governments. An e-form is the electronic version of its corresponding paper form. E-forms have some benefits over paper forms including eliminating the cost of printing, storing, and distributing pre-printed forms. In addition, e-forms can be filled out faster because the programming associated with them can automatically format, calculate, look up, and validate information for the user.

To cope with the requirements of different local governments, different e-forms need to be developed. To capture the commonality and variability of e-forms a domain analysis to e-forms is required that will result in a domain model. One of the common techniques for domain modeling is feature modeling, which has been extensively used in domain engineering [6]. Hereby, a feature model is a result of a domain analysis process in which the common and variant properties of a domain are elicited and modeled. In addition, the feature model identifies the constraints on the legal combinations of features.

In principle, feature models can be used for manually implementing e-forms. However, to support reuse and productivity, Excellence aimed for automatic generation of e-forms. Automatic generation has been broadly addressed in model-driven software development (MDSD). One of the basic pillars in MDSD is defined by model transformations and likewise several useful approaches have been proposed in this context [6][1].

Within the context of the industrial research project we have applied model-driven engineering techniques for the automatic generation of e-forms (electronic forms) from feature models. This project has shown that feature modeling is an effective means not only to model the domain of e-forms but also to support the automatic generation in a model-driven engineering process. Besides this observation the results of our study also presents an additional insight and lessons learned regarding model transformation practices in general. In particular it appeared that for defining e-forms offline static single generation is less suitable. This is because the specific e-forms depend on the user input and the retrieved data from the data administration services. In this paper, we show three different approaches for generation with increasing complexity: (1) off line model transformation without interaction (2) model transformation with initial interaction (3) model transformation with run-time interaction. We report on our experiences and lessons learned and propose a systematic approach for defining model transformations that is based on an interactive paradigm.

The outline of the paper is structured as follows: In section 2 we provide the problem statement together with the case study on e-forms for local governments. In section 3 we provide the necessary preliminaries for supporting the explanation of the approach in section 4. Section 5 describes the need for enhancing the model-transformation pattern with interaction. Section 6 elaborates on the generation process by considering workflow aspects. Section 7 provides the evaluation. Section 8 presents the related work, and finally section 9 presents the conclusions.

## 2. Problem Statement

In this section we will illustrate the problems that we address using the real industrial case of Excellence. Different cities in The Netherlands also provide e-government services and likewise different e-forms are used for each local government. Using e-forms on the internet site of the local governments, citizens can perform requests such as electronic filing of taxes, licenses and permits, payment of utilities, and general information about government services.

Figure 1 shows different e-form examples for *notification of movement* that are used by two different cities, Enschede and Deventer in The Netherlands. The citizens can access the local government gateways to request e-government services. Note that although the same service is implemented the e-forms are different with respect to the required information and the order in which this is asked, and the presentation of the fields and the e-form in general.

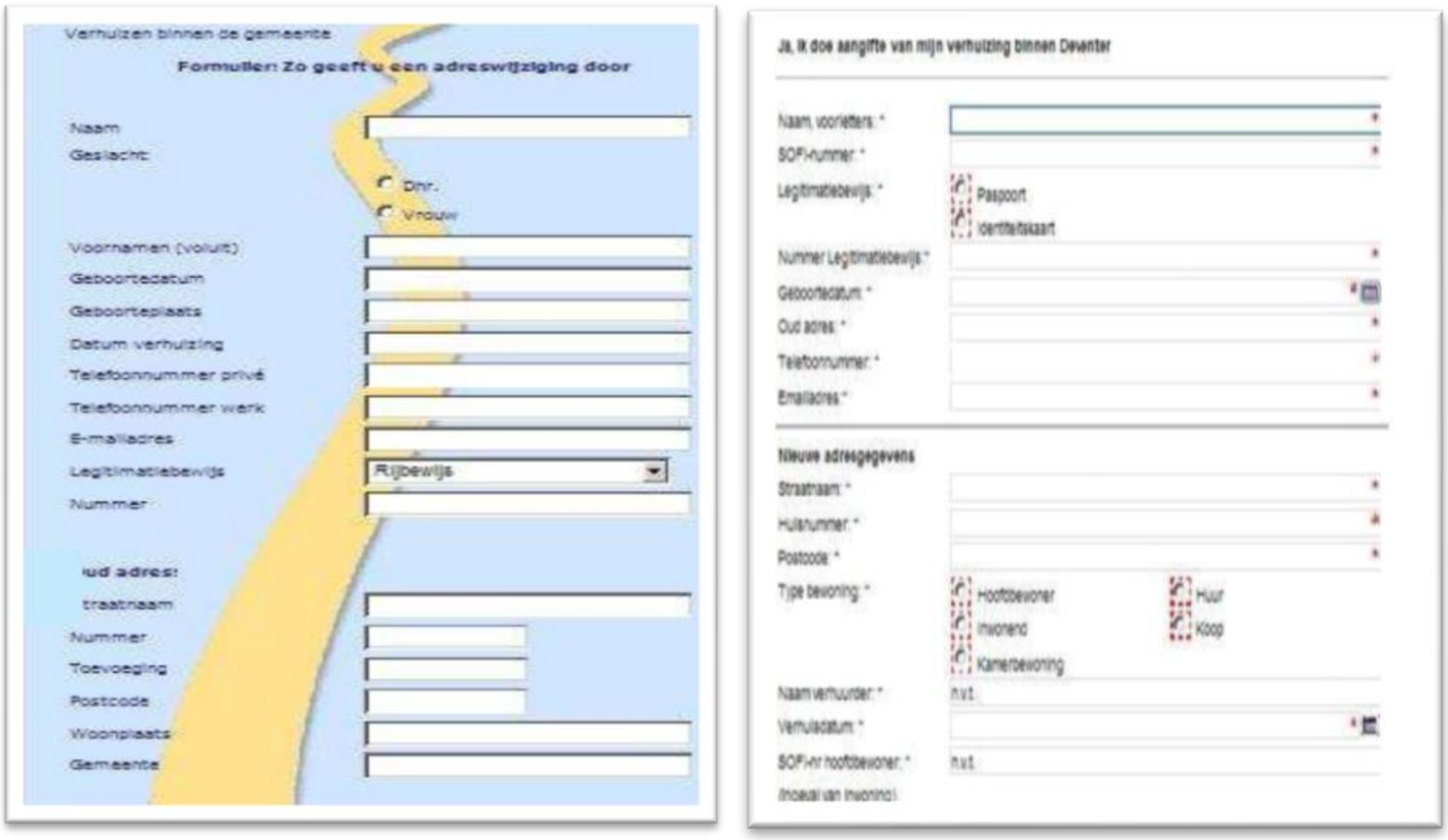

**Figure 1**. Example e-government interface of two different examples of local governments for supporting e-services

The initial deployment view for supporting e-forms by Excellence is depicted in Figure 2. E-forms are manually developed by developers and remain at a web server of a local government. End-users (citizens) can access these web pages through internet browsers. E-forms are usually defined over multiple web pages. Once a user logs in to the system, the user can select a number of services offered by the local government, such as the notification of movement. A middleware layer, defined by the so-called *MidOffice* server includes functions to access personal data of the registered users in the local government, which is stored in one or more back office systems, the *Data Administration*.

Based on the selected product and the user, the information about the user is requested from the common administration through *MidOffice*. The unknown fields are filled out by the user. After the citizen enters the last field the system needs to generate a report and submit the request to the government clerk. An important advantage of the *MidOffice* is the loose coupling between the interfaces (presentation) and the back offices (data). Different back office system can be accessed by different web browsers. The sclient web pages only communicate through the *MidOffice* which is responsible for the communication and distribution logic.

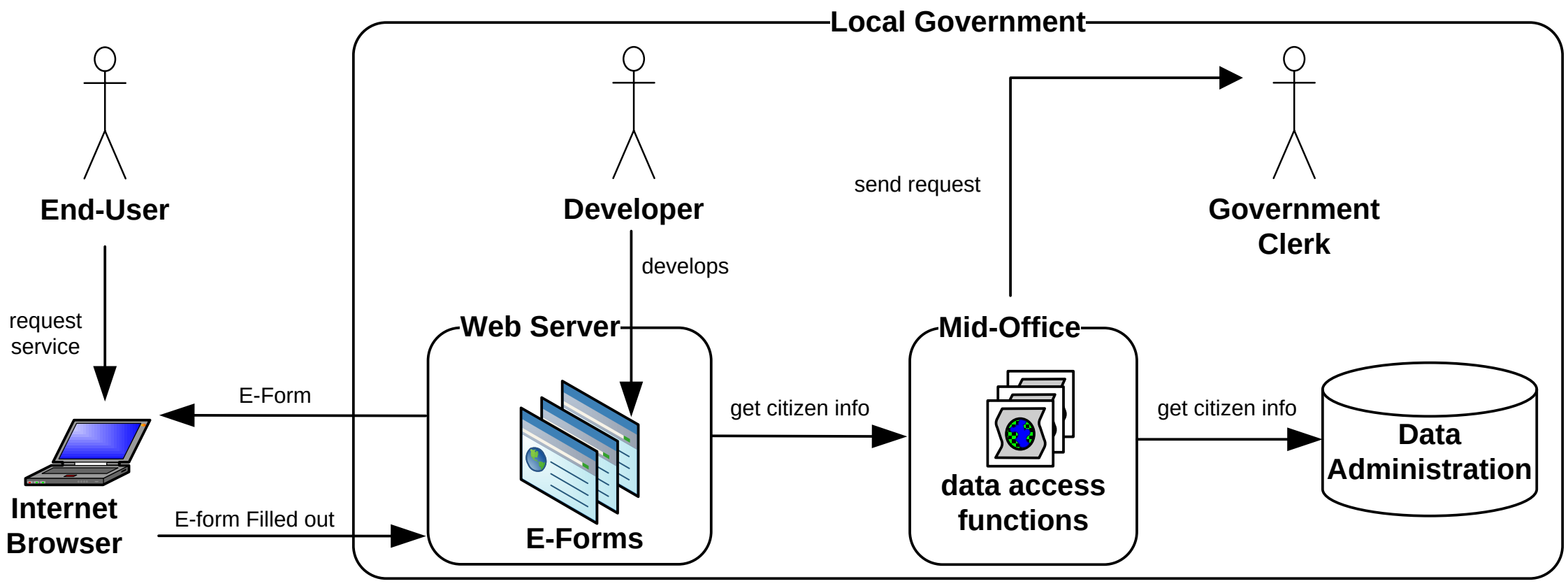

**Figure 2**. Architecture with Manually Developed E-Forms

In the initial version of the system, e-forms were manually implemented and deployed on the webserver of local governments. A number of obstacles with this manual, static development solution can be identified. We distinguish the concerns for three different stakeholders:

**Developer:**

- *Lack of reuse of e-forms*

First of all, even when we are dealing with the same kind of service, such as notification of movement, different local governments might require different kind of e-forms. The differences might be in the required type of data, the presentation form or the control flow i.e. the order in which the data is presented to the citizen. Although the e-forms share much commonality, the lack of systematic variability management requires that for each different local government an e-form needs to be implemented from scratch. On its turn lack of reuse results in low productivity and longer time to develop the e-forms.

- *Maintenance of e-forms*

Even after deployment of the e-forms on the web servers, based on earlier practical experiences, updates might be required to the implemented e-forms in due time. Unfortunately, the maintenance of the web pages including the e-forms is not trivial and again requires changes to the requested data, the presentation form or the control flow.

**End-User:**

- *Misaligned e-forms for different users*

The exact fields that need to be asked to the user for a given service may depend on the input of the user. It appears that very often some fields in the e-forms can only be known when a citizen is filling out the e-form. Hence, the values for these fields in the e-form can actually only be known at run-time. Because of this limitation usually the complete e-form is provided to the user, which complicates the process of filling out the form by the citizen. The e-form would be easier if only the required information is presented using multiple web pages that are presented at the right time.

- *Time to fill out e-forms takes too long*

Due to misaligned fields that do not necessarily align with the situation of the citizen, the time to fill out the e-forms might take unnecessarily long.

**Local Government:**

- *Citizen Empowerment*

An important goal of e-government is to enhance active participation of the citizens in the society. The ease of accessing the gateway of the e-government, downloading, filling out and automatic sending of e-forms to the corresponding e-government clerks will ease the task of the citizen and as such support more active participation.

- *Citizen Satisfaction*

For the local government it is important to enhance citizen satisfaction as much as possible. In case the citizens feel that they cannot access the channels to the local government entities, this will negatively impact the citizen satisfaction.

## 3. Preliminaries

In order to support the explanation of the approach as described in the next section we will discuss the basic modeling approaches first.

### 3.1 Feature Modeling

To cope with the requirements of different local governments, there is a need for modeling the commonality and variability of E-forms. *Feature modeling* is a popular, systematic approach for describing variabilities and commonalities of systems in the context of software product lines [32][43][30]. *Feature models* were first introduced in the Feature-Oriented Domain Analysis (FODA) method by Kang in 1990 [22]. A *feature model* is a model that defines features and their dependencies. *Feature* is defined as a characteristic of the system that is observable to the user. Feature models are usually represented in feature diagram (or tables). A *feature diagram* is a tree with the root representing a concept (e.g., a software system), and its descendent nodes are features. Relationships between a parent feature and its child features (or subfeatures) are categorized as follows: *(i) Mandatory* – if a child feature is required. *(ii) Optional* – if a child feature is defined as optional, that is, it can be included or not. *(iii) Alternative* (xor) – only one of the sub-features could be included in every product that contains the parent (iv) Or – at least one of the sub-features must be selected. A *feature configuration* is a set of features which describes a product member of a product family. A *feature constraint* further restricts the possible selections of features to define configurations. The most common feature constraints are: (i) *requires* – if the selection of feature A in a product implies the selection of feature B. (ii) *excludes* – if feature A is selected then feature B cannot be part of the same product.

Different extensions have been provided to this basic feature modeling approach. The cardinality-based feature modeling approach [9] associates each feature with a feature cardinality that specifies how many clones of the feature are allowed in a specific configuration. Further, features can be organized in feature groups, which also have group cardinality. This feature group cardinality defines the minimum and maximum number of group members that can be selected. A feature can have further an *attribute type*, which indicates that an attribute value can be specified during configuration. Usually one attribute per feature is allowed. The attribute type can be a basic type, such as *String* or *Integer*. Figure 3 shows the adopted notation in the CBFM approach.

| Symbol | Explanation |
|---|---|
| f | Solitary feature with cardinality [1..1] (mandatory) |
| f | Solitary feature with cardinality [0..1] (optional) |
| f [n..m] | Solitary feature with cardinality [n..m] |
| f ▶ | Feature model reference f |
| f(value T) | Feature f with attribute of Type T and value of *value* |
|  | Feature group with cardinality [1..1], i.e. xor |
|  | Feature group with cardinality [1..k], i.e. or |
| [n..m] | Feature group with cardinality n..m |

**Figure 3**. Cardinality-Based Feature Modeling Notation; adapted from [9]

Different e-forms are implemented for different local governments but besides of the variations one can easily observe commonality of requested data. To model the domain for a given service we can define a *family feature model*. Figure 4 shows, for example, a feature diagram for a service of a local government, which is the notification of moving. This feature model defines the space of requested data that can be implemented on different e-forms. To put it differently, the feature diagram represents an intermediate representation for the space of e-forms.

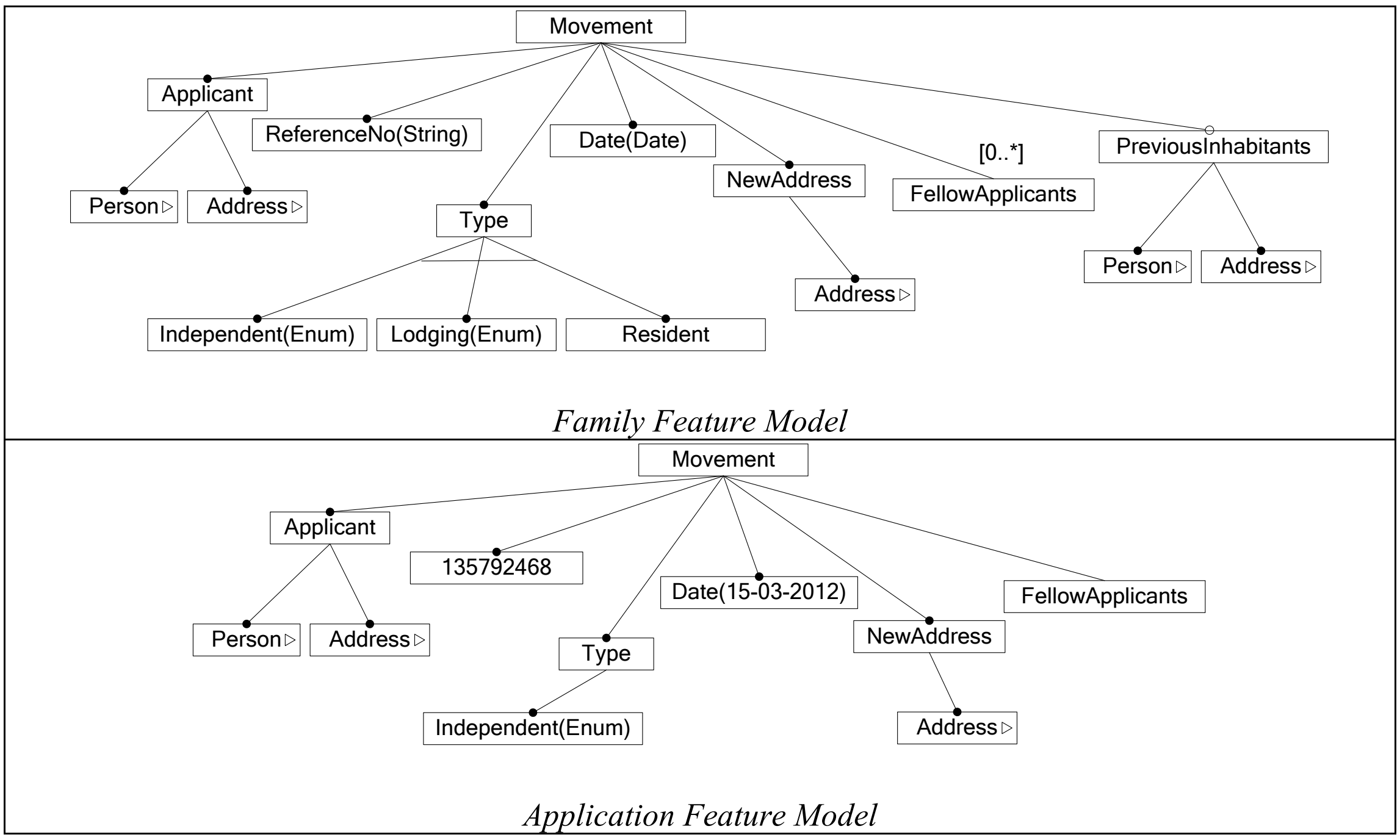

**Figure 4**. Family feature diagram of service *notification of movement* and corresponding example application feature model

Different instantiations of the feature diagram indicate different definitions of e-forms. An example instantiation of the family feature diagram is shown in the lower part of Figure 4. Based on the application feature model the corresponding e-form can be implemented.

### 3.2 Model-Based UI Design

User interface is a very important part in software development. It appears that an average of 48% of the code of applications is devoted to user interface, and about 50% of the implementation time is required for implementing the user interface [19][28]. Historically, information systems construction and user interface design have traditionally been undertaken using different methods and techniques. Model-based user interface design (MBUID) aims to integrate user interface modeling and model-based design and support the automatic generation of user interfaces. Automatic generation of user interfaces improves the quality of developed interfaces, supports maintainability and reduces the cost of development. MBUID approaches are supported by Model-based user interface development environments (MBUIDE) which is a tool suite that supports the design and automatic creation of interfaces using high-level declarative models. Usually each MBUIDE defines its own set of models to describe the interface but a common structure can be observed as well. In general the MBUID approaches adopt the following types of models to support the design and generation of user interfaces [7][17][19][24][44][39]:

- *Task and Domain Model* describes the tasks that the user can perform. The domain model is a high level representation of the objects and their associated functions in a given model.
- *Abstract User Interface (AUI)* is an abstract model of an interface that is independent from the underlying concrete model. Usually, the AUI is expressed in terms of presentation units that are independent of the specific type of interactors as well as the modality of interactions (e.g. graphical, vocal).
- *Concrete User Interface (CUI) Model* is an expression of the UI in terms of concrete interactors that are specific for the selected type of platform. Concrete interactors are the abstractions of UI components that are generally included in toolkits.
- *Final User Interface (FUI)* is the source code implementing the user interface in a programming language or mark-up language. The source code can be interpreted or compiled, and rendered.

To define a user interface first the task model is described, which is mapped to an AUI, then to CUI and finally the code. The mappings define either semi-automatic or automatic transformations. Note that in MBUID the principles as advocated by the Model Driven Architecture (MDA) [24] has been adopted. To support quality factors such as portability, interoperability and reuse, MDA explicitly separates the functionality from platform specific concerns and provides Computation Independent Models (CIMs), Platform Independent Models (PIMs), Platform Specific Models (PSMs) and the code (model). The development of a system in MDA starts with defining the CIM, which is mapped to a PIM, and through a series of transformations gradually the platform specific properties are included through the platform specific models, eventually resulting in the final code [42]. In this respect, the Task/Domain Model corresponds to the CIM, the AUI to the PIM, the CUI to PSM and the final user interface to the code level in MDA. Both the MBUID and the MDA stack are shown in Figure 5.

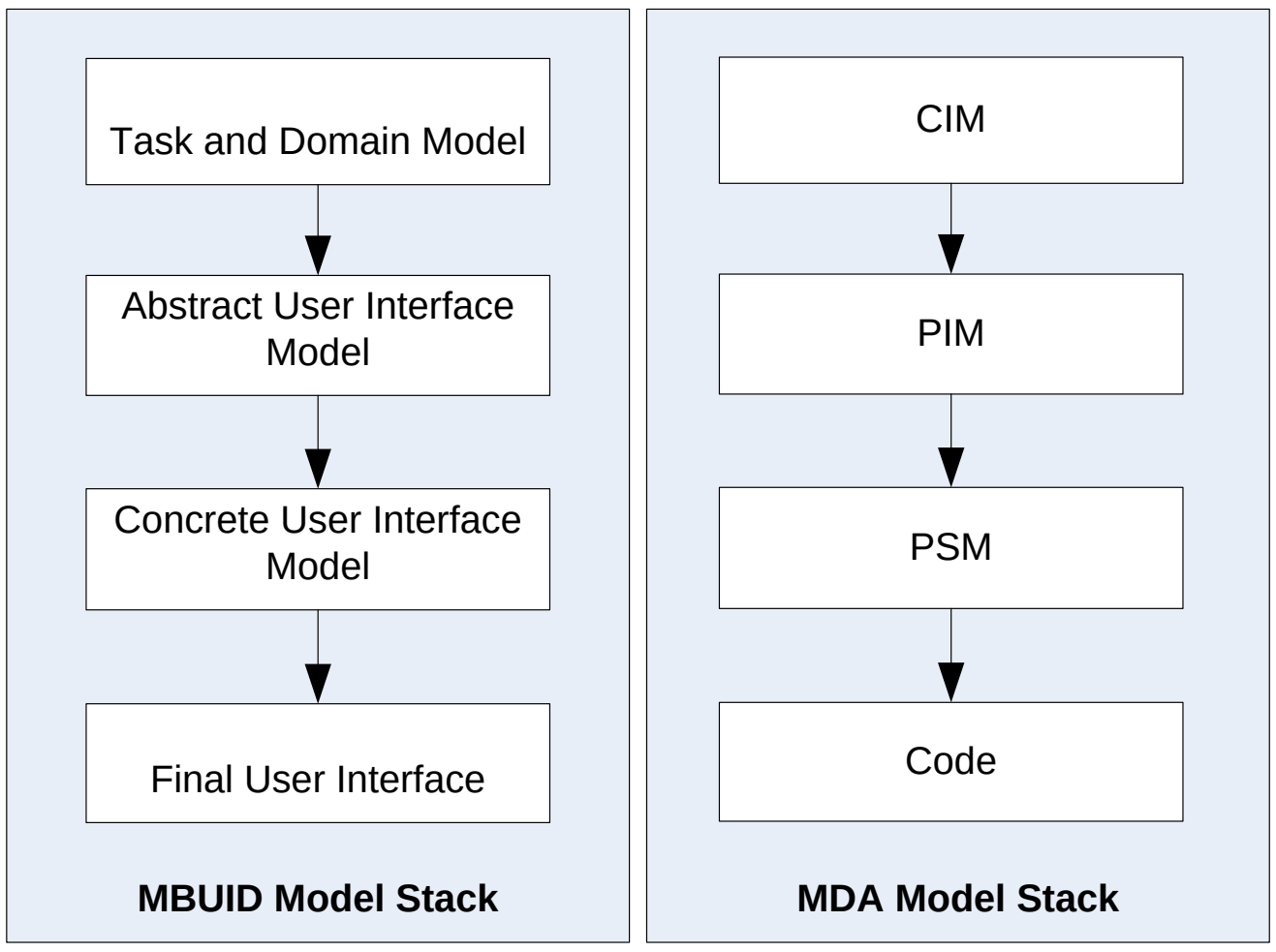

**Figure 5.** Four levels of abstraction that is adopted in Model-Based User Interface Design Approaches and relation to MDA Model Hierarchy [7]

# 4. Approach

## 4.1 Overall Process

To support the automation of e-form generators we have adopted the approach as shown in Figure 6. The process is an instantiation of the product line engineering process that is usually defined in the literature [45]. The SPLE process consists of life cycle processes of *domain engineering* and *application engineering*. The domain engineering process is responsible for establishing the reusable platform and thus for defining the commonality and the variability of the product line [34]. The platform consists of all types of software artefacts (requirements, design, realization, tests, etc.). In the application engineering process the applications of the product line are built by reusing the artefacts and exploiting the product line variability as defined in the domain engineering process. The application design process takes as input the application requirements and by reusing the product line architecture as defined in the domain engineering phase, the application architecture is developed and finally the application is implemented reusing the assets from the platform. In essence, the domain that we consider is a product line of e-form generators. The approach that we present in Figure 6 is an integration of the product line engineering process and the MBUID model stack as shown in Figure 5. Hereby the Domain Model is represented using feature models. We distinguish among a family feature model that is defined in the domain engineering process, and application feature model that is defined in the application engineering process. Based on the family feature model the abstract user interface model is defined in the domain engineering process. The family feature model is used to define the application feature model in the application engineering phase. An example of the family feature model and application feature model has been presented before in Figure 4. Each family feature model defines a particular service, such as *Notification of Movement* such as shown in Figure 4. The application feature model represents the instantiation of the family feature model for a particular citizen. Once the service is instantiated the concrete user interface model is defined for a particular local government institution. The final step is the generation of the code from the concrete user interface model.

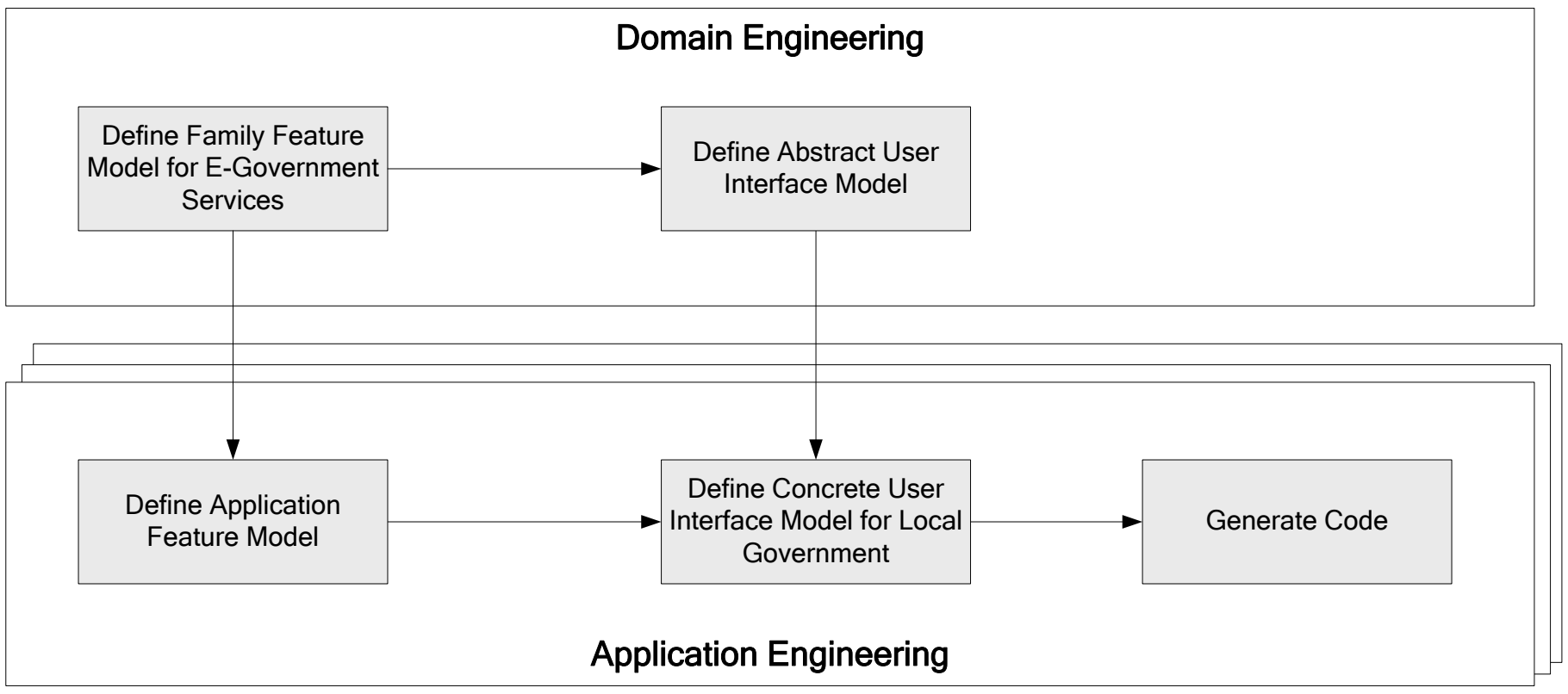

**Figure 6.** Approach for generating E-forms within the context of e-government

### 4.2 Model Transformations

Figure 6 does not consider the issue whether the transitions among the models are manual or automatic. In fact, the depicted process is general and also represents the current case for defining user interfaces using abstract declarative models. As stated before to cope with the problems of the manual development of user interfaces, for the particular case study of Excellence Company it was decided to automate the process in Figure 6 as much as possible. For implementing the required generator we adopt model-transformation patterns as defined in the model-driven development [5][10][18][24][34][41]. Independent of the Excellence case we have first defined the general generator model which is defined as a chain of model transformations as illustrated in Figure 7. Here, each model transformation gets as input a source model and provides as output a target model. In general a distinction is made between model-to-model transformations (M2M) and model-to-text transformations (M2T). In the M2M case both the input and the output models conform to their metamodel, in the M2T case only the source model conforms to its metamodel whereas the target model is usually text or code that does not have a metamodel. As such, the generator in Figure 7 consists of three M2M and one M2T transformations.

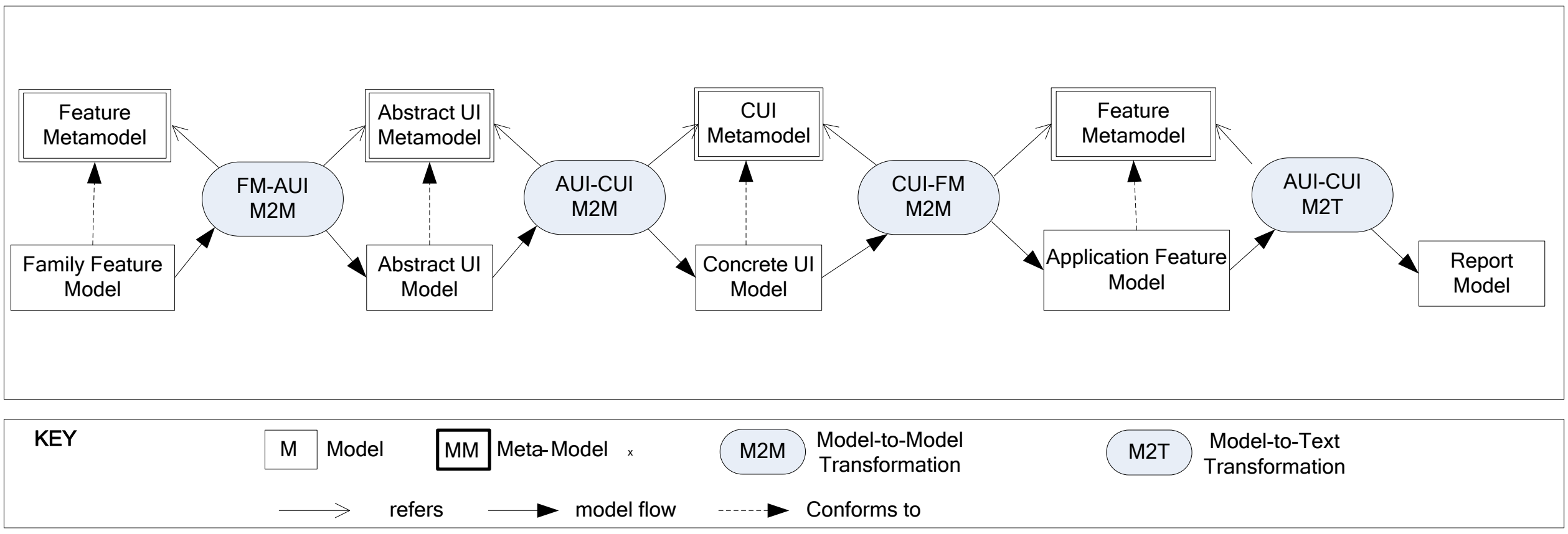

**Figure 7**. Chain of model transformations as defined in the generator of Figure 8

The generator includes the following four transformations:

*(1) Family feature model to AUI model* ($T_{FM\text{-}AUI}$) – The family feature model of the service will be used to generate the UI model representing the e-form.

*(2) AUI to CUI model* ($T_{AUI\text{-}CUI}$) – A CUI is generated based on the AUI and the platform requirements of the local government. In case AUI is missing then the previous transformation will include the mapping from FM to CUI directly, and likewise the mapping from AUI to CUI will be omitted.

*(3) CUI model to application feature model* ($T_{CUI\text{-}FM}$) – Based on the entered values by the user in the UI model, the application feature model will be generated.

*(4) Feature model to Report model* ($T_{FM\text{-}R}$) – After all the fields in the e-form are filled out, and the final feature model is generated, a report will be generated.

Each of these transformations can be further specified. For the case of Excellence two decisions were made that customized the model transformation process of Figure 7. The first decision is the instantiation of the family feature model through web pages (i.e. UI). That is, the user input is used to instantiate an application feature model from the family feature. The second decision was the direct transformation of feature models to CUI, that is, no AUI was defined. The transformation from AUI to CUI is usually needed when so-called *elastic interfaces* are required [7], that is, interfaces that can be adapted to different devices (desktop, laptop, mobile phone, smart pads etc.). In the case of Excellence the concern for elastic interfaces was not considered as an urgent issue.

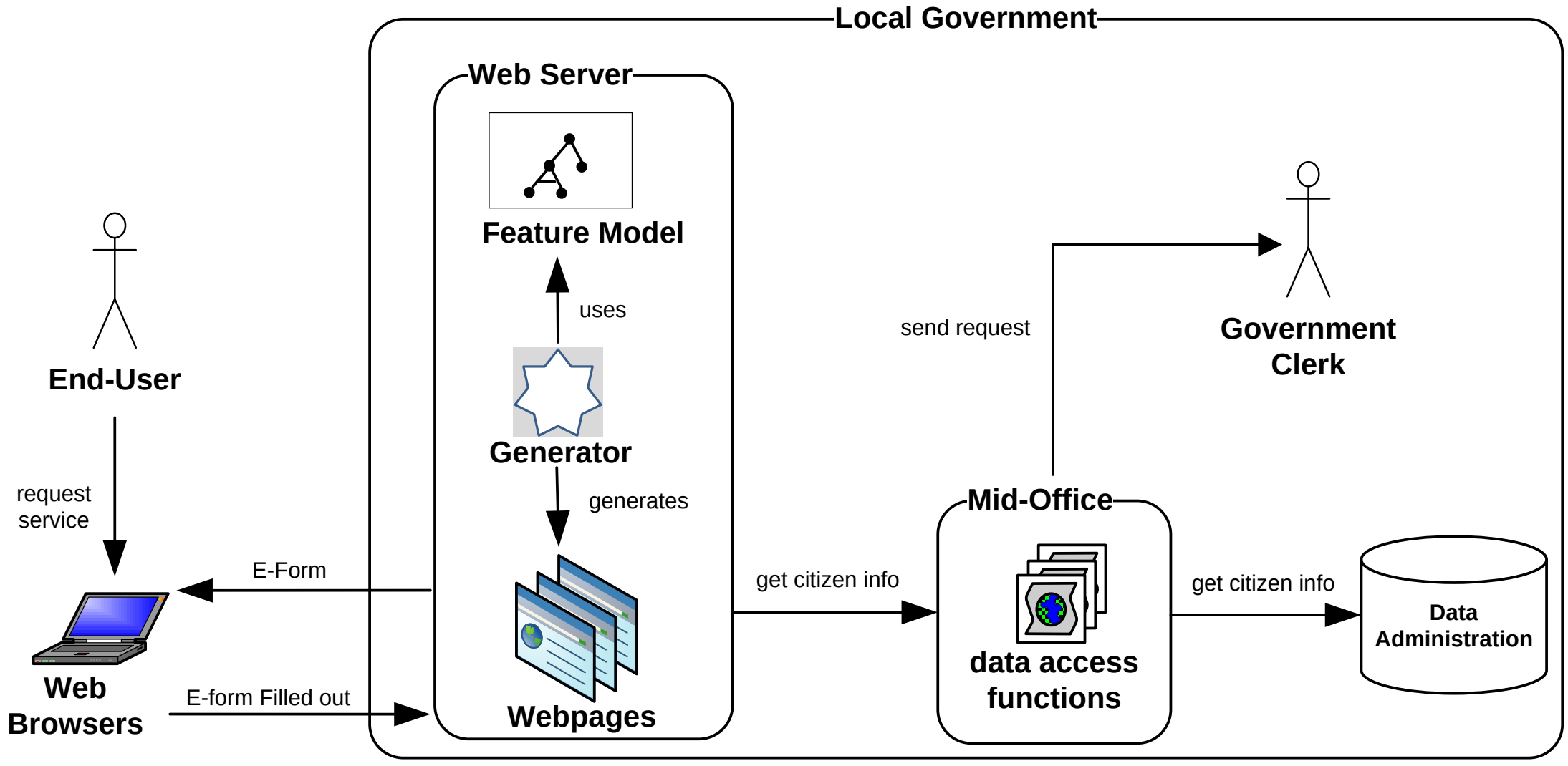

**Figure 8**. E-form automatic generation

Given the general process in Figure 6, the model transformation chain in Figure 7 and the decisions for instantiating feature models through UI, and the exclusion of AUI, the manual development process as defined in Figure 2 is now replaced with Figure 8. This model now includes a generator for automatically generating e-forms based on the given feature models and the user interface model. Hereby, upon the connection with the gateway of a local government, the citizen selects a particular product service (e.g. movement). The generator retrieves the family feature model for the product service and generates the required web pages with the required user interfaces. The citizen enters the required values as defined in the generated UIs, and if necessary the generator can access details from the data administration.

After all the values are entered the final instantiated feature model is defined which is then generated to a report that is sent to the government clerk.

### 4.3 Required Metamodels

For defining the model transformations in the generator of Figure 8 we need to define the necessary metamodels and the model transformation rules. In the model transformation chain we can distinguish among the feature metamodel and user interface metamodel. The feature metamodel which is shown in Figure 9 is based on the metamodel of the cardinality based feature modeling approach [9], which has been explained before in section 3.1. For our own purposes we have slightly modified the original more extensive feature metamodel.

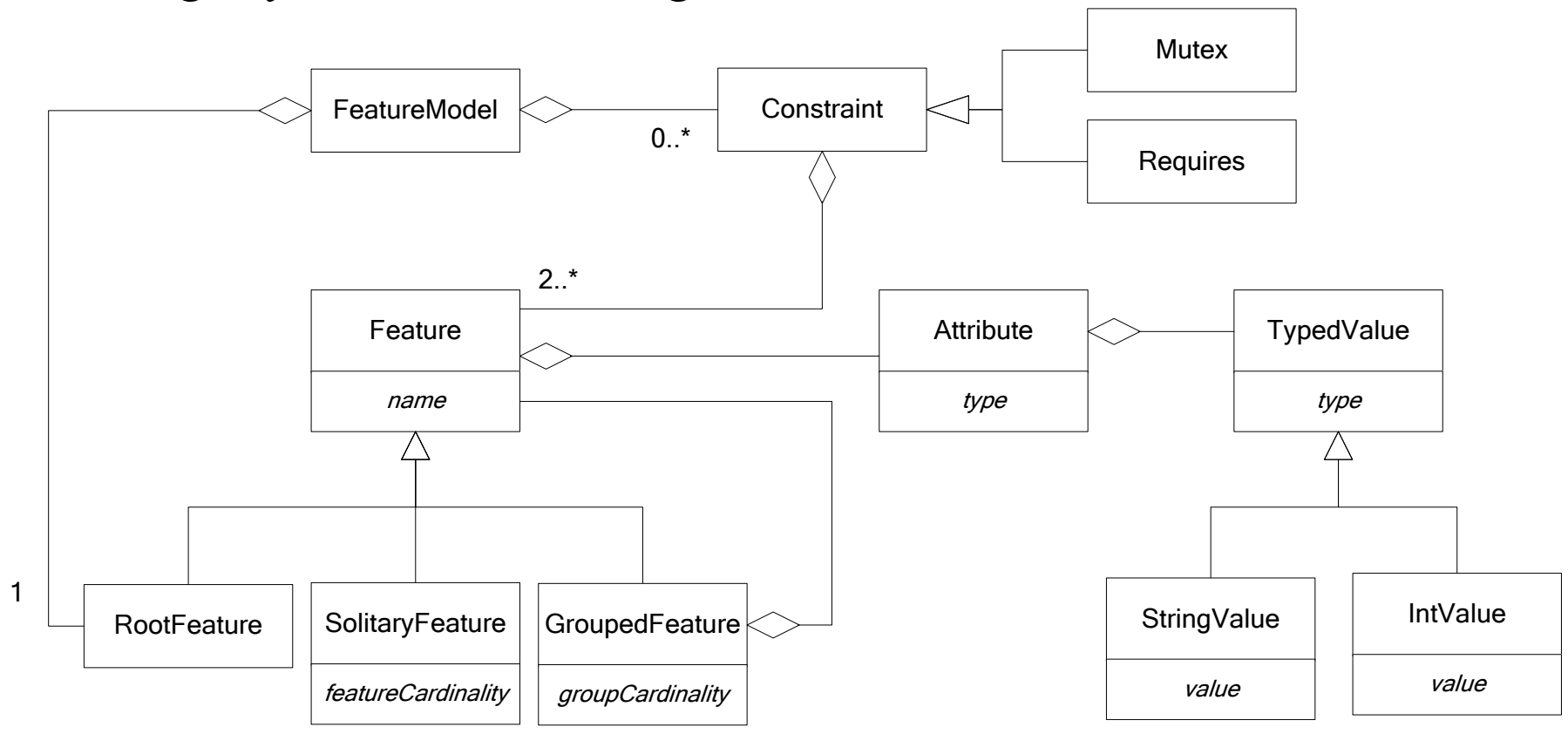

**Figure 9**. Adopted Feature Metamodel (Adapted from [9][35])

Since for the case of Excellence it was decided not to use AUI we only needed to define the metamodel for CUI. The CUI metamodel as shown in Figure 10 has been adapted from different sources including [17][23][14]. Each Web Application consists of a number of pages that include Widgets. Widgets can be either Input, Output, Navigation or Layout.

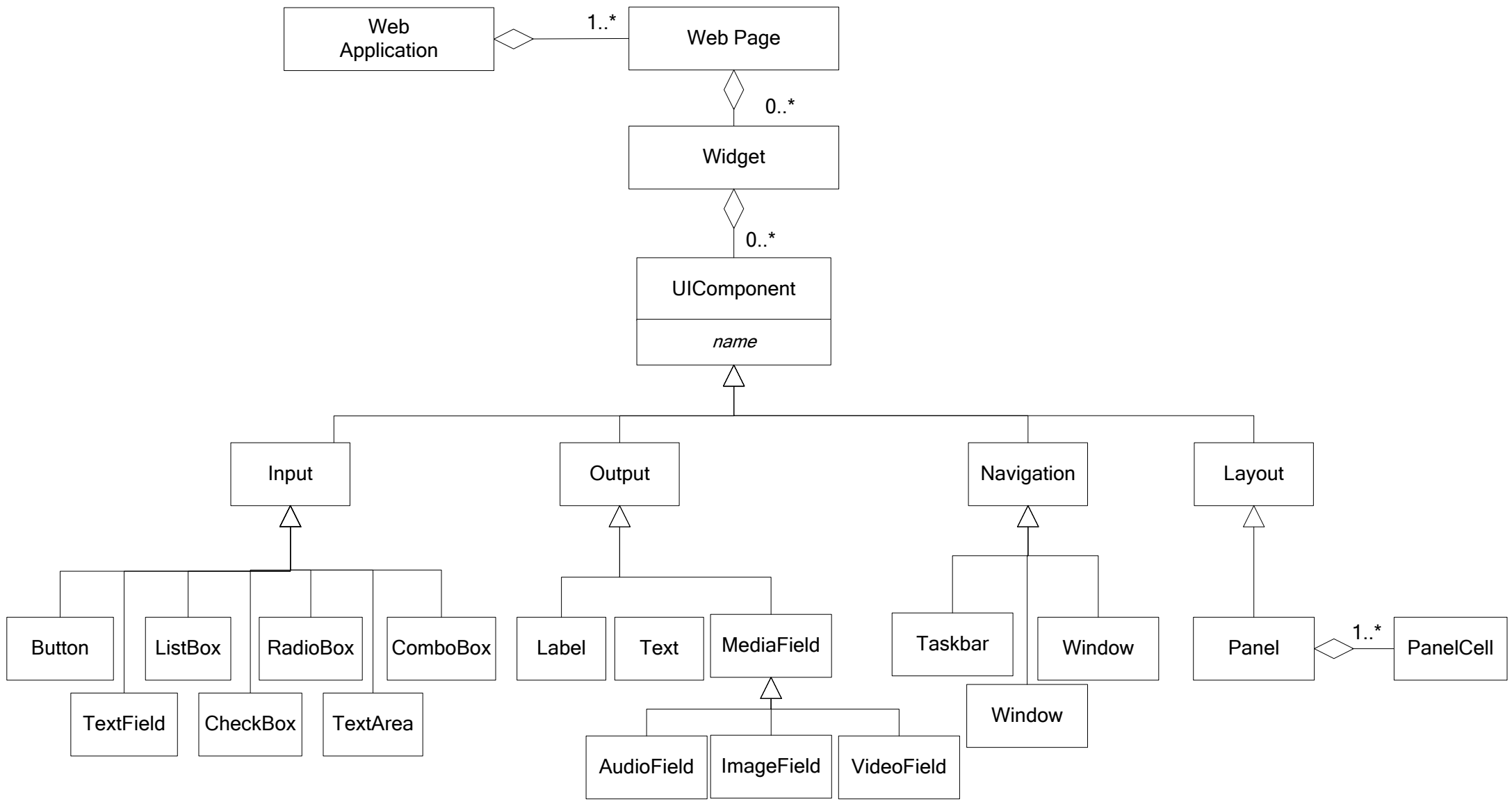

**Figure 10**. UI Metamodel (Adapted from [17][23][14])

### 4.4 Transformation Rules

Excellence is extensively working with XML technology. Likewise, for implementing the metamodels we have chosen to represent these in XML schemas. Accordingly, the required transformation rules in the model transformation have been implemented using XSLT, a language that can be easily used for transforming XML documents to other XML documents. The following transformations were implemented in the Excellence context:

*(1) Feature model to UI model* ($T_{FM\text{-}UI}$) – The feature model of the service will be used to generate the UI model representing the e-form. The feature model could be either a family feature model or application feature model. For the transformation this does not matter much.

*(2) UI model* to Feature model ($T_{UI\text{-}FM}$) – This is the reverse of the previous transformation. This transformation is needed to define the representation of the feature model (in XML) that is used to generate a report.

*(3) Feature model to Report model* ($T_{FM\text{-}R}$) – After all the fields in the e-form are filled out, and the final feature model is generated, a report will be generated.

The transformations are defined using a set of transformation rules that are implemented using XSLT. Each transformation follows the IF <Condition> THEN <Action> format. Here, <Condition> refers to the conditions that must be valid to execute the transformation rule. Typically, a condition checks the state of a feature in the feature model of the service. Example conditions are checking whether the feature is a solitary node or group node, whether it has an attribute or not, whether feature is mandatory, optional or alternative etc. The second part of the rule, <Action>, defines the particular action to realize the transformation. Depending of the three above transformations are executed this indicates the update of a feature model, UI model or report.

**Table 1.** Selected rules for FM-UI transformation

| Tranformation Rule ID | Entry Conditions | Action |
|---|---|---|
| TR1 | First Element | Create body of the webpage including *html, head, title,* and form |
| TR2 | Element = SolitaryFeature and *child* element = Description | Create element *label* |
| TR3 | Element = SolitaryFeature and *child* element = Attribute and and attribute type = string, integer, float | Create element *input* with attributes *type(text), name* and *value* |
| TR4 | Element = FeatureGroup and number of *child* element is more than one | Create element label; Create element *input* with attributes *type(radio), name* and *value* |

Each of these rules in Table 1 is implemented using XSLT. For example the implementation of rule TR4 is shown in Figure 11.

```
<xsl:for-each select="fm:FeatureGroup">
   <br/>
   <xsl:choose>
      <xsl:when test="count(fm:GroupedFeature)=1"></xsl:when>
      <xsl:otherwise>
         <label style="width: 150px">
            <xsl:value-of select="fm:Annotation/fm:Description/@fm:value"/>
         </label>
         <xsl:for-each select="fm:GroupedFeature">
            <input>
               <xsl:attribute name="type">radio</xsl:attribute>
               <xsl:attribute name="name">
                  <xsl:for-each select= "ancestor::*">
                     <xsl:text>/</xsl:text>
                     <xsl:value-of select="@fm:value"/>
                  </xsl:for-each>
               </xsl:attribute>
               <xsl:attribute name="value">
                  <xsl:value-of select="@fm:value"/>
               </xsl:attribute>
               <xsl:for-each select="fm:Attribute/fm:String/
               fm:StringProperties/fm:StringValue">
                  <xsl:value-of select="@fm:value" />
               </xsl:for-each>
            </input>
         </xsl:for-each>
         <br/>
      </xsl:otherwise>
   </xsl:choose><xsl:apply-templates />
</xsl:for-each>
```

**Figure 11**. Implementation of rule TR4 in the transformation FM to UI

The definition of transformation rules and their implementation has been defined in a similar way for the other transformations.

## 5. Model-Transformation with Interaction

So far we have seen the benefit of the integration of feature modeling and model-driven transformation techniques. Whilst feature models capture the commonality and variability of the data, model-transformations can be used to automate the development process. The overall model-driven process in section 3 also largely supports the goals for automated development of e-forms. Unfortunately, the model-driven transformation approaches as they are applied in practice do not seem to solve all the problems that we have encountered for the case. The main reason for this is that the transformation process does not take into account interaction with the user and the data administration. The transformation process takes as input a feature model to generate an e-form in which all the required fields are presented to the end-user. The end-user needs to fill out all the requested data and the e-form can only be completed if all the information is entered. Once the e-form is complete a report is generated and the service request is submitted for handling.

This manual implementation of e-forms with only weak interaction with citizen and/or back offices is to some extent doable but certainly not cost effective. When filling out the e-form, interaction with the Data Administration might be required to retrieve data to speed up the process or to complete the e-form. Obviously, explicitly addressing interaction in model transformations is here a key issue. Model-transformation based on interaction is also important to better align the E-forms for the particular citizen and the data that is entered during the processing of the e-form. The notion of model transformation is of central importance to model-driven software development. Unfortunately, in current model-driven development practices interaction is not explicitly addressed. In Figure 12, we propose a slightly enhanced model transformation pattern that enables interaction for the transformation. Hereby, the transformation engine does not only execute the defined transformation but also interacts with a so-called *Interaction Source.* This interaction might require additional values and as can have a direct impact on the transformation result.

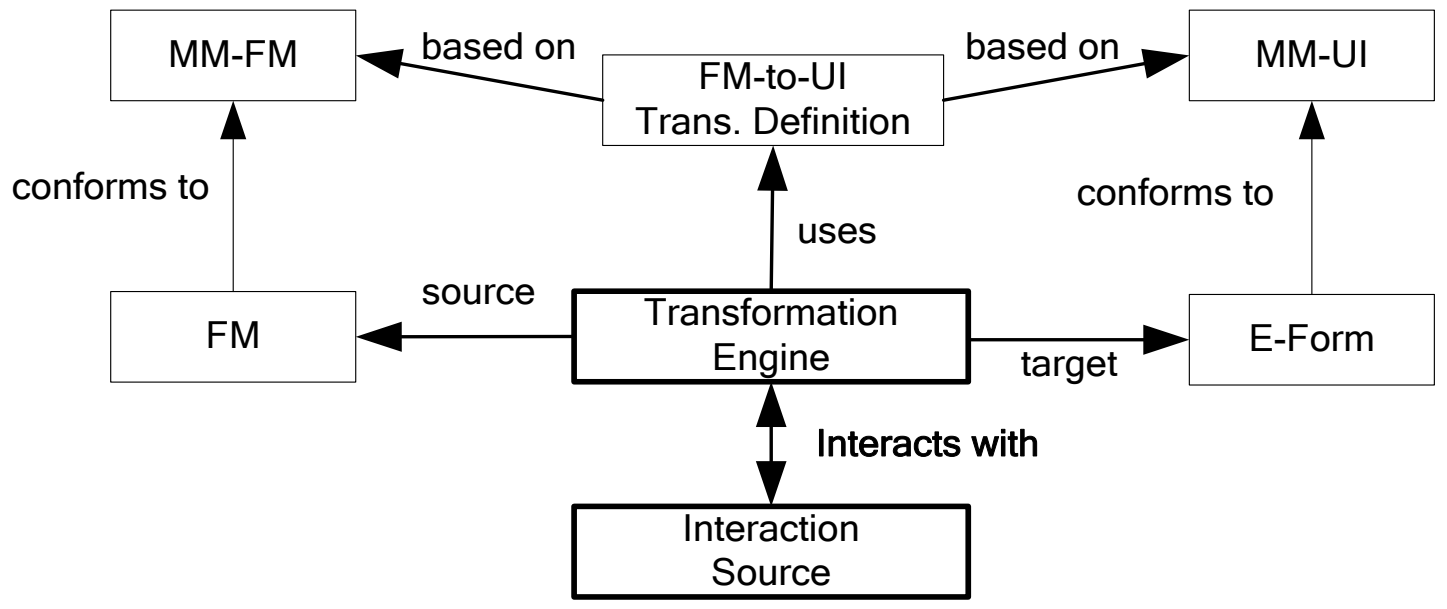

**Figure 12**. Enhanced Model Transformation Pattern with Interaction

Using the model in Figure 12 we have defined two different types of generators based on the frequency and time of interaction:

1. *Generator with single, initial interaction*. This generator is similar to the generator that adopts the conventional transformation pattern, but allows for initial interaction with the data administration server to retrieve the values for the fields that can already be defined in the e-form. The generator adopts the model as defined in Figure 12 but with only a one-pass, single interaction.

2. *Generator with multiple, run-time interaction*. This generator complements the previous generator by allowing interaction with user and data administration during run-time. For this a number of functions of data administration can be invoked to speed up the e-form completion process. The generator adopts the model as defined in Figure 12 in which interactions between *Transformation Engine* and *Interaction Source* takes place until the target model is created.

In the following we describe each of these generators in more detail.

### 5.1 Generator with single, initial interaction

This generator makes use of an initial call to the data administration (i.e. Interaction Source). After the authentication process and selection of a particular service the system can already retrieve some information about the citizen and the selected product and instantiate part of the feature diagram. As such, the time to fill out the form, as well as the chance for incomplete forms will be partially reduced. Compared to the generation process without interaction, this generation process actually includes one more transformation pattern. This is the transformation from a source feature model to another target feature model. The order of steps that is taken in this generation process is shown in Figure 13a.

After the authentication of the user and the selection of the corresponding product service, the family feature model is loaded. Then the initial call to data administration is made to retrieve personal details, and the application feature model is based on the input retrieved from the data administration. The final step includes the generation of the complete e-form based on application feature model.

### 5.2 Run-time Interaction

The first generator without interaction already solves the automation problem of e-forms. By defining transformations e-forms can be automatically generated. The second generator with initial interaction with data administration retrieves the necessary data that could be filled out. As such the e-form completion process time is reduced. However, both generators generate one complete web page in which all the fields are shown. Unfortunately, this is not always suitable since the generation of the specific fields in the e-form also depends on the data that is entered by the user, or the data is retrieved from the data administration, at run-time.

The *generator with multiple, run-time interaction* allows run-time interaction with the user and data administration. In this way, the e-form is generated incrementally dependent on the input of the end-user. This means that the instantiation of the family feature diagram is not done after authentication process but at any time during completing the e-form. Also multiple web pages including part of the e-form are generated.

The interaction process for this generator is shown in Figure 13b. After the authentication process, the family feature model is retrieved and the first fields are defined. Then follows a cycle of interaction with user and data administration in which the application feature model and likewise the corresponding e-form is specialized. Once the e-form is complete a report is generated and the request is submitted.  In essence the transformation process is similar to the generator without interaction. The main difference is that now the feature model is specialized multiple times and during the e-form completion process. Obviously, multiple model transformations are required to complete the process. In fact, this process also follows the idea of staged configuration of feature models as explained in [9].

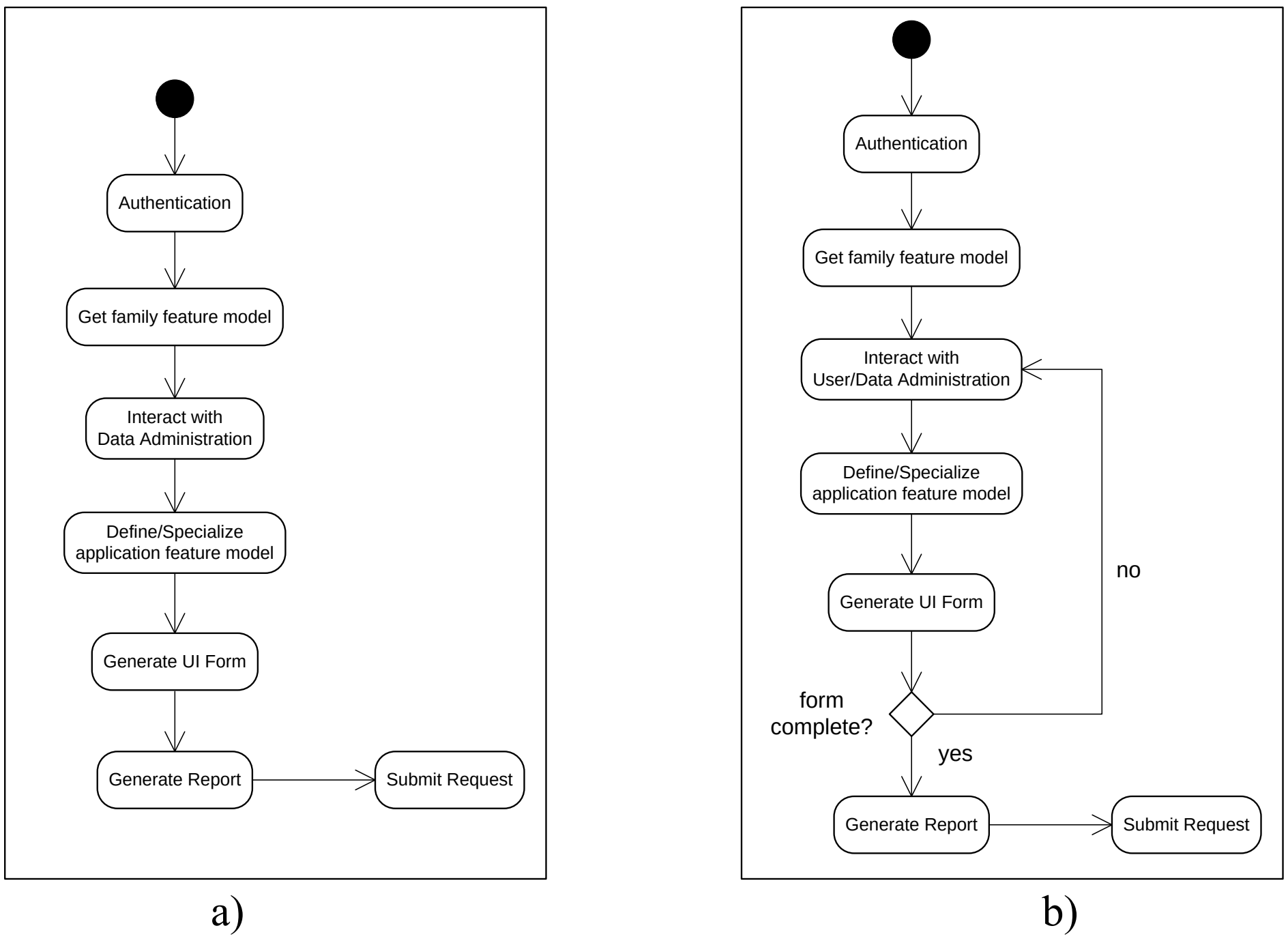

**Figure 13**. Activity diagram for a) Generation Process with initial interaction b) generation process with run-time interaction

## 6. Optimizing Workflow

When interaction with the data administration is supported the functions for data administration are accessed. Many different functions might be accessed given an application feature model. For example, the invocation of the function *getPersonDetails* can define the values for name, address, and id of the citizen. Further, each invocation of a function might result in the definition of the values of different fields.

In essence the aim is to optimize the e-form completion process and therefore the functions need to be preferably invoked in the order in which the maximum set of values in the e-form can be determined. The latter means that the number of fields that the citizen needs to enter is optimally reduced.

It appears thus that we need to address the workflow explicitly to optimize the generation process. In the first generator without interaction no data administration function was called at all. In the second generator with initial interaction only one call was made to the data administration. As such the workflow concern was not considered in these two generators. In the generator with run-time interaction the workflow concern is explicitly as illustrated in Figure 14. Hereby, besides of the *Interaction Source*, *Transformation Engine* uses *Workflow Engine*. *Workflow Engine* accesses the set of *Workflow Functions* that can be invoked, and defines the order in which these functions are processed. As such based on the state of the e-form (and the application feature model) a decision needs to be made which functions of the data administration need to be called. This is defined by *Workflow Strategy* in Figure 14.

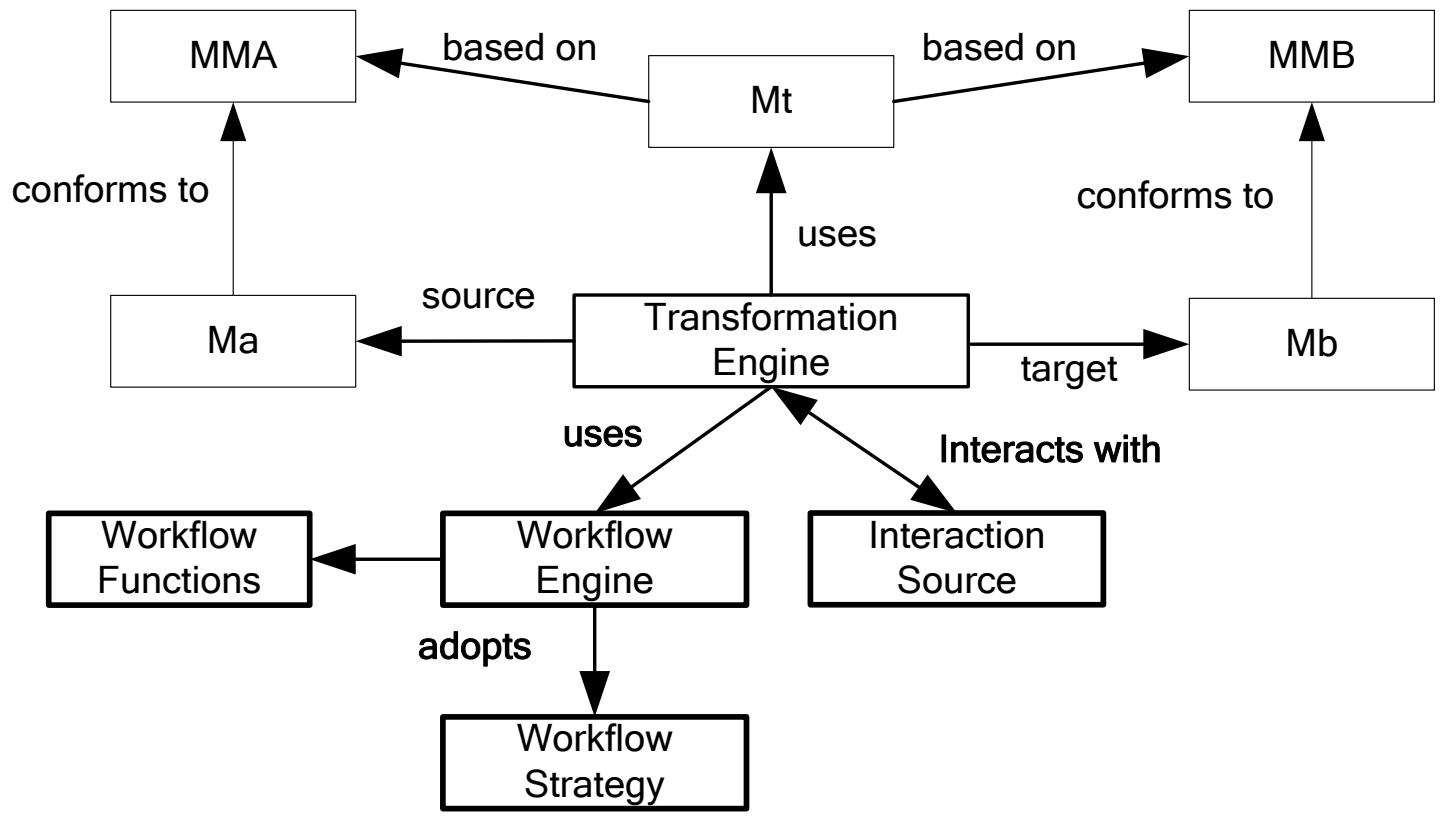

**Figure 14**. Enhanced Model Transformation Pattern with Interaction and Workflow Engine

Different strategies can be adopted by the workflow engine. We have adopted a fixed, strategy which aims to optimize the number of model transformations needed. The workflow definition is defined as depicted in Figure 15. Hereby we first check whether mandatory features have been defined in the feature model. These are then first processed, that is an e-form is generated with these fields, and data input from the user is processed resulting in a new feature diagram. The following step is to select features that are related to functions in the data administration. The final step is the generation of optional features. Once all the fields have been entered the report is generated. In fact this is quite a simple workflow strategy and can be optimized in different ways. For example, we could prioritize the functions that result in more input from data administration; we could define the optimal path of these functions, etc. The full integration of strategy selection and optimization has been reserved for the future work.

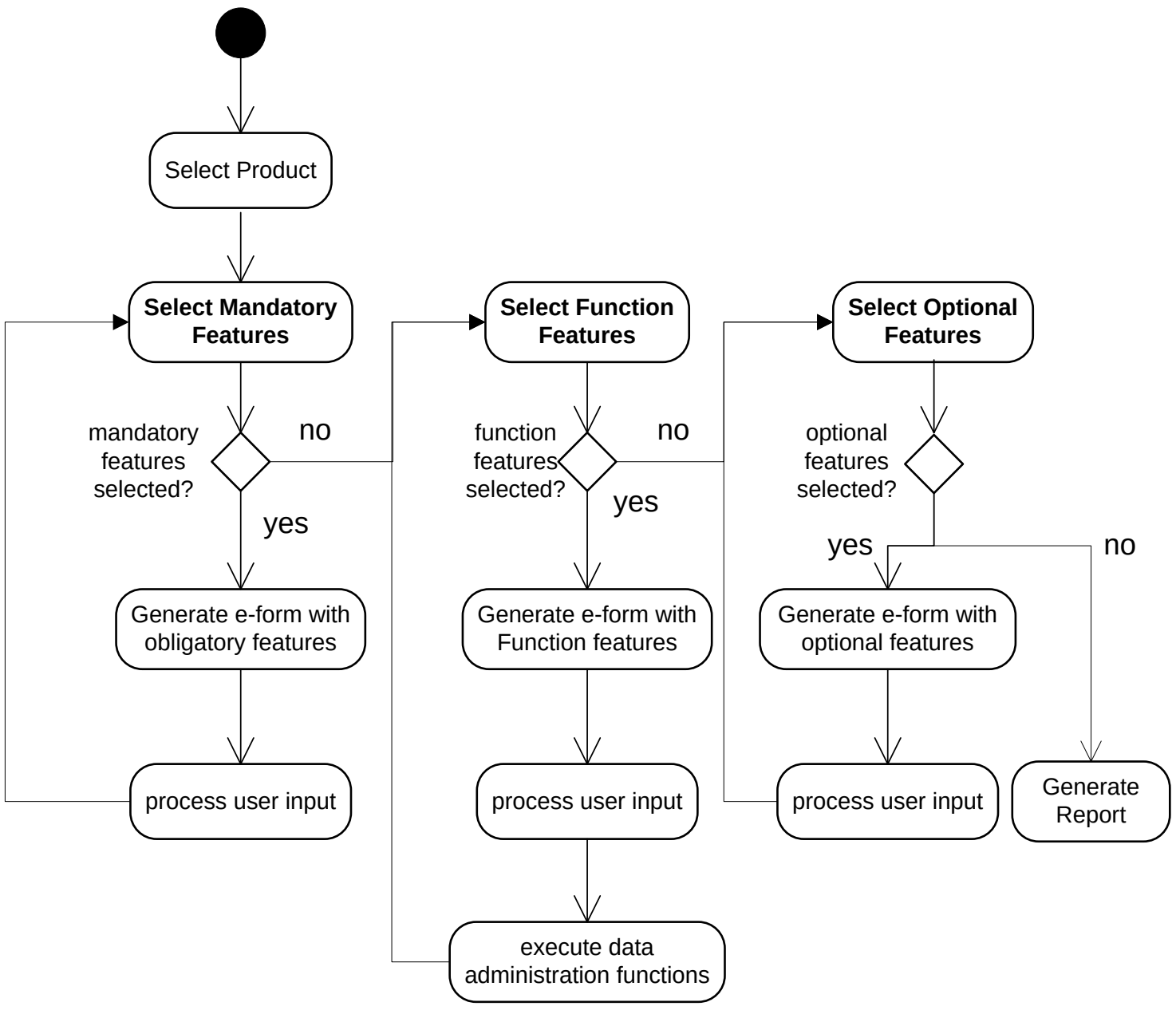

**Figure 15**. Adopted workflow in the interaction-based e-form generator

# 7. Evaluation

In this paper we have discussed four different alternatives for developing e-forms: (1) manual (2) automatic e-form generation without interaction (2) automatic e-form generation with initial interaction, and (4) automatic e-form generation with run-time workflow. In the following sub-sections we provide a qualitative evaluation and the quantitative evaluations for the important metrics.

## 7.1 Qualitative Evaluation

The different alternatives have been evaluated with respect to the stakeholders' concerns which were described using the corresponding metrics. The results of the evaluation are shown in Table 2. For each stakeholder the key concerns have been defined using the scale --, -, +-, +, ++ (low to high). As the table shows the alternative of automatic e-form generation with run-time workflow has been considered as the most optimal approach.

Obviously, the manual form is the least preferable from all the three perspectives of developer, end-user and local government. For the developer it is hard to develop, reuse and maintain the forms. For the end-user the fixed (set of) manual forms is often not easy to fill out. For the local government it will be hard to process e-forms and also the concerns for citizen satisfaction and citizen empowerment will be impeded. Of course the manual form can already be supported by defining separate feature models before production of the e-forms. The feature models could nicely capture the commonality and variability of the features for a given product service. A feature model that is defined for a given product service can now be easily reused for multiple local governments. Hence, regarding the overall goal of reuse seemed to be achieved from this perspective. Because of the elegant separation of data and presentation, the variability of the presentation format can now also be more easily addressed. However, without the automation of the e-forms the adoption of feature models is still limited.

The MBUID approaches were evaluated positively with respect to the manual form development. Hereby, the data (feature model), the presentation (AUI and CUI) and control flow are nicely separated. The data is represented by feature models, the presentation and control flow are separately carried out by the generator that defines the mapping from abstract user interface to concrete user interface and lastly the final user interface. In case a local government requires a different presentation format, the style sheet that is applied in the generator is updated. This process is relatively easy to do and can be reused for all the product services in the corresponding local government context. As such, for the developer the reuse is enhanced, the time-to-develop is shortened and the maintenance of e-forms is supported. The time-to-develop criterion for generator with run-time workflow is evaluated less positively because of the complexities related to control flow modeling and implementation. The control flow is separate from both the data and presentation and is addressed by a separate workflow engine. But this part seemed to be the hardest in defining the generator because of the difficulty for defining a generic optimal workflow definition for all product services. In general each product service might require a different optimal workflow definition. Also the workflow optimization is dependent on the specific user interactions at run-time. In our study we have defined one fixed control flow that first checks the mandatory features, and then checks the list of functions to find the function that results in the highest set of values in the next step. The time performance of this strategy did not cause any problems for the given case and as such it can be reused for different local governments.

**Table 2. Evaluation of the Manual Forms and the different E-Form Generation Strategies**

| Stakeholder | Concern | Approach | | | |
|---|---|---|---|---|---|
| | | **Manual** | **Automatic No Interaction** | **Initial Interaction** | **Run-time with workflow** |
| **Developer** | Reuse of E-Forms | -- | ++ | ++ | + |
| | Time-to-Develop E-forms | -- | ++ | ++ | +- |
| | Maintenance of E-Forms | -- | ++ | ++ | + |
| **End-User** | Dedicated Questions | -- | +- | + | ++ |
| | Save Time to fill out | -- | +- | + | ++ |
| **Local Government** | Fast Service Handling | -- | +- | + | ++ |
| | Cost Reduction | - | + | + | |
| | Citizen Satisfaction | - | + | + | ++ |
| | Citizen Empowerment | - | + | + | ++ |

### 7.2 Quantitative Evaluation

For Excellence one of the key concerns is the time to develop the e-forms with the selected adopted approach. To provide a more quantitative evaluation Excellence Company has undertaken a case study in which the manual development of e-forms was compared to the generator with workflow. The time-to-develop includes both the time to develop the elements in an UI form and the time to implement the functions to access the database to support the run-time interaction with the generator.

For the evaluation we have defined two different product services that needed to be implemented, *excerpt request* and *felling permit*. *Excerpt request* includes the request for a formal report about the citizen's registration details in the local government. *Felling Permit* includes the request to fell a tree in a public or private area by the citizen. To implement these services in the earlier, non-generator based approach the developers had to implement the e-forms manually. As stated before the e-forms integrated the data, presentation and the control flow. In the generator-based approach the developer basically needs to (1) define the family feature model for the product service, and (2) implement the functions for the corresponding features. The generator for the e-form, together with the interaction and workflow engineer can be reused.

For the evaluation process we have provided the scenarios to two different teams for developing the e-forms. One developer team had to develop the e-forms for the scenarios using the existing approach. The other developer used the generator based approach to develop the e-forms. Part of the representation for the feature models for the two development scenarios is given in Figure 16.

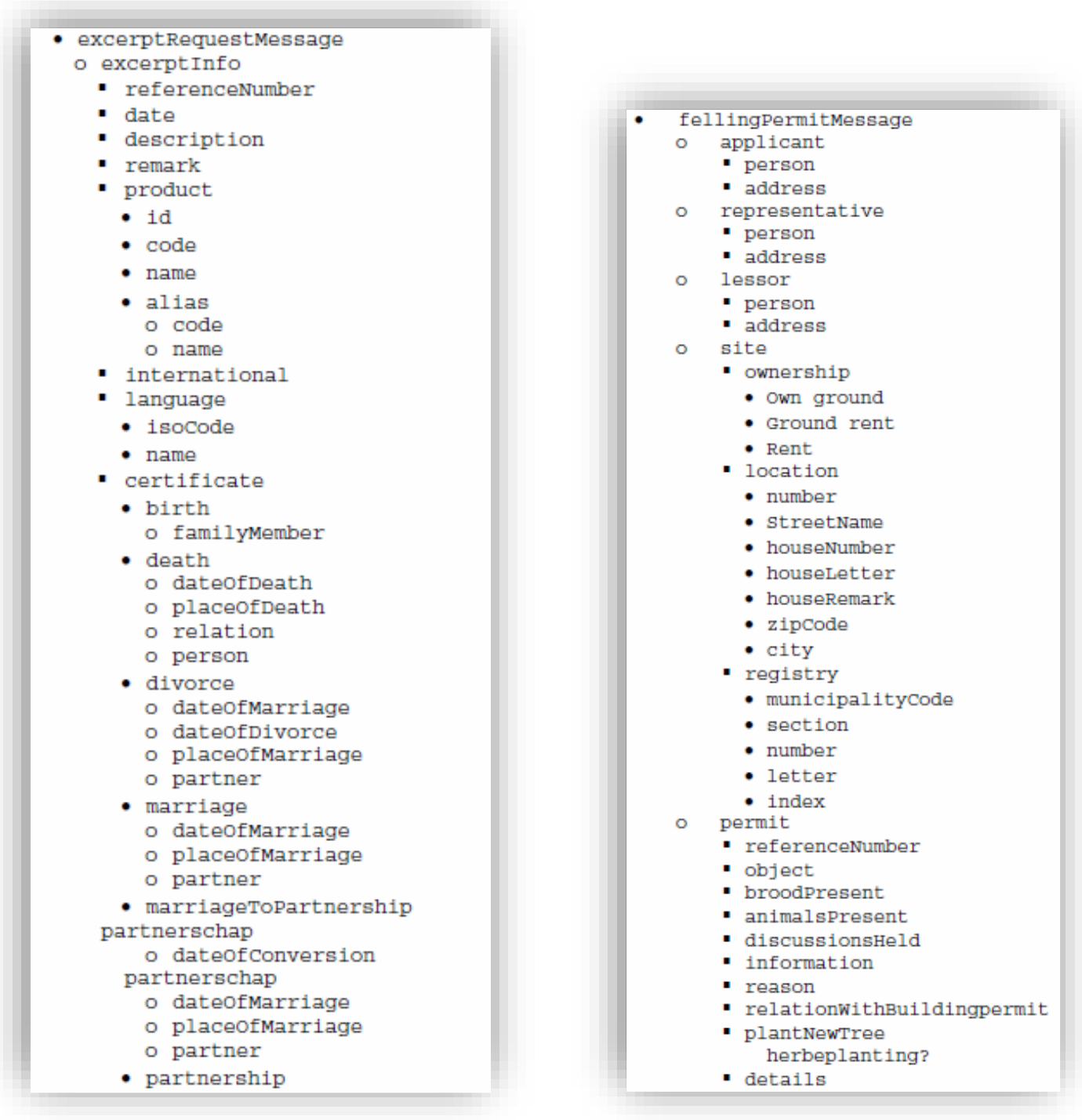

**Figure 16**. Evaluation scenarios for developing e-forms using conventional approach or generation-based approach

Given these scenarios to two different teams the development time has been measured. The result of the evaluation is shown in Table 3.

**Table 3.** Results of the Evaluation

| **Approach** | **Scenario 1 – Excerpt Request** | **Scenario 2 – Felling Permit** |
|---|---|---|
| *Initial Approach* | 3 hrs and 52 min | 3 hrs and 18 min |
| *Generation –based approach* | 2 hrs and 50 min | 2 hrs and 15 min |

For these two scenarios it appeared that the generator-based approach performed at least as good as the current approach. In fact, in the generation-based approach the e-forms were developed about one hour faster than in the initial approach.

Together with the qualitative analysis of the approach the Excellence has decided to adopt the generation-based approach for developing e-forms. It is expected that the adoption of the generation process will include some upfront investment (in particular education in feature-driven development, model-driven software development, etc) but for the midterm and long-term a return-on-investment is expected.

## 8. Related Work

The notion of UI design has been extensively discussed in the HCI domain [38][28]. Many different approaches have been proposed to enhance the quality of user interfaces. A large part of the research is focused on manual design of user interfaces. Hereby, UI designers start from textual requirements and apply the provided UI design principles, but they do not systematically consider domain models such as feature modeling. In the HCI design we can also observe several approaches which aim to fully automate the user interfaces. These seem to however only effective for very specific applications domains and fail to provide a user interface with sufficient quality [6][32][33]. In the Excellence project the UI designs were also defined manually. Later on we have investigated the automation of the user interfaces. Our experiences also revealed that this is indeed very hard. Although, initially we tried to provide a fully automated system, we soon realized that this is not practical with respect to the frequent adaptability requirement and the very diverse requirements of the local governments. Therefore, in addition to reuse we have also allowed editing the feature models or UI models directly, when this deemed to be necessary.

The Model-Based UI XG Final Report [28] provides a summary of the work and results obtained by the Model-Based User Interfaces Incubator Group (MBUI-XG). The goal of MBUI-XG is to support new generation Web authoring tools and define mechanisms for tailoring Web applications for different user preferences, device capabilities and environments. The report provides the result of an evaluated research on MBUI as a framework for authoring Web applications and with a view to proposing work on related standards. Further, the report provides a nice overview of several state-of-the art approaches for task modeling, AUI modeling and CUI modeling.

Pleuss et al. [32] combine and adapt several concepts from Model-based User Interface Development (MBUID) and integrate them into the SPL product derivation process. Similar to our general approach they also distinguish among abstract user interface (AUI) and concrete user interface (CUI). The AUI is defined in the domain engineering phase and the CUI is calculated during the application engineering. The final UI is derived using semi automatic-approaches from MBUID. The authors indicate that some elements like the links between UI elements and application can be fully automatically generated while the visual presentation can be both generated automatically and customized by the user.

In this paper we have focused on extending model-transformations [37][5] with interaction behavior that use feature models to select the required fields in E-forms. An interesting work that is directly related to model transformation is the paper by Czarnecki and Helsen who use feature models to classify model-transformation approaches [10]. They model transformation according different feature dimensions. One important dimension represents the *rule application control* which has two aspects: *location determination* and *scheduling*. *Location determination* is the strategy for determining the model locations to which transformation rules are applied. The *location determination* strategy includes three features *deterministic, non-deterministic* or *interactive. Scheduling* determines the order in which transformation rules are executed. Using these dimensions of *location determination* and *Scheduling* we can characterize our approach as a model transformation approach in which the location determination and rule selection is interactive. Another taxonomy of model transformations is provided by Mens and van Gorp [24]. The taxonomy in this paper is essentially multi-dimensional and is more targeted towards tools, techniques and formalisms supporting the activity of model transformation. Both papers are very useful in depicting the current space of model transformation approaches in MDSD. We should, however note that the notion of interaction is not addressed in detail in the classifications. One obvious reason for this is of course that interactive model transformations are not yet in the mainstream approaches of MDSD.

In [16] France and Rumpe provide a research roadmap of model-driven development of complex software systems. They indicate that an important difficulty of developing complex software is caused due the wide conceptual gap between the problem and the “implementation domains of discourse”. The current research in MDE aims to reduce this gap through the explicit use of models and systematic transformation of problem-level abstractions to software implementations. One important observation of France and Rumpe is that the current research in MDE mainly focuses on producing implementation and deployment artefacts from detailed design models. They argue that for building complex systems that dynamically adapt to changes in their operating environments requires runtime monitoring and execution of software. Hence, they encourage the use of run-time models. The special issue on *models@runtime* in IEEE Computer [4] also shows the need and development of the ideas in this trend. In our approach, an important indication of this requirement forms the need for run-time interaction of model-transformation engines. We could model commonality and variability using feature models, and define automatic transformation to develop E-Forms. However, without the runtime interaction in transformation it is very hard to develop e-forms that follow the run-time input of the end-user.

Debnath et al. [2] provide a feature modeling to categorize e-government systems. In addition they provide an initial approach for integrating formal specification with feature models to provide consistency rules. Peristeras et al. [31] provide a review on model-driven initiatives and efforts to achieve e-government interoperability. The identified initiatives are grouped into categories based on the owner, scope and modelling perspective of each project. Most of the initiatives seem to be based on XML Schemas. Also in our approach all the models that we have used for implemented using XML, while the transformations were primarily based on XSLT.

## 9. Conclusions

In this paper we have discussed our experiences with using MBUID to a local e-government project. Our experience related to different important domains including e-government, software product line engineering, user interface design, and model-driven development. Only the synthesis of these approaches helped us to solve the hard problems in a real industrial and commercial setting.

In the context of e-government domain we have focused on the generation of E-forms using model-driven techniques. To cope with the commonality and variability of E-forms domain modeling using feature models was essential. On the other hand, the need for automation required the integration of feature modeling with model-driven approaches. Using the conventional model transformation pattern we have identified different transformation patterns including feature model to UI, UI to feature model, and feature model to report. All these transformations supported the automation process of e-forms and as such improved reuse and productivity. In addition we have pinpointed the necessity for interaction in generating e-forms. This is because the e-form is not only defined by the selected service but also defined by the entered answers in the e-form or the retrieved information from the data administration. To cope with this issue, model transformations could not remain static and/or offline but had to be integrated in the run-time e-form completion process. Based on the input at important steps in the e-form completion process the application feature model was regenerated and in accordance with this the e-form updated. It also appeared that hereby the order in which the functions of the data administration are accessed, i.e. the workflow, has an impact on the e-form completion process. In alignment with this issue, we have shortly discussed the notion of workflow concern.

We think that the lessons that we have derived from the considered project should be considered from a general and broader perspective. In particular the issue of interaction in the model-transformation process is a topic that needs further investigation. From the UI perspective we think that our experiences are also of value both from the theoretic and practical perspective.

## Acknowledgments

We would like to thank Mehmet Aksit from the University of Twente, and Anton Boerma and Richard Scholten from Excellence  for earlier support and discussions about this work.